\documentclass[sigconf,screen]{acmart}

\AtBeginDocument{%
  }

\usepackage[export]{adjustbox}
\usepackage{bbding}
\usepackage{array}
\usepackage{graphicx}
\usepackage{multirow}
\usepackage{booktabs}
\usepackage{xcolor}
\usepackage{makecell}
\usepackage{float}
\usepackage{tikz}
\usepackage{subcaption}
\usetikzlibrary{positioning,shapes.geometric}
\usepackage{balance}

\definecolor{crq}{HTML}{ededed}
\definecolor{crq2}{HTML}{dee2e6}
\tikzstyle{rqbox}  = [draw=crq2, fill=crq, very thick, rectangle,
                       rounded corners, inner sep=10pt, inner ysep=12pt]
\tikzstyle{titlerq}= [fill=crq2, draw=crq2, rounded corners, inner sep=4pt]

\setcopyright{cc}
\setcctype{by}
\acmDOI{10.1145/3832783.3834554}
\acmYear{2026}
\copyrightyear{2026}
\acmISBN{979-8-4007-2882-2/2026/10}
\acmConference[ASE '26]{Proceedings of the 41st IEEE/ACM International Conference on Automated Software Engineering}{October 12--16, 2026}{Munich, Germany}
\acmBooktitle{Proceedings of the 41st IEEE/ACM International Conference on Automated Software Engineering (ASE '26), October 12--16, 2026, Munich, Germany}
\acmSubmissionID{ase26nier-p79-p}
\received{2026-05-13}
\received[accepted]{2026-07-02}

\begin{document}

\title{Do Not Copy/Paste: Soft Barriers for Copying in AI-Assisted Programming}

\author{Iyiola E. Olatunji}
\correspondingauthor
\orcid{0000-0002-0391-9202}
\affiliation{%
  \institution{University of Luxembourg}
  \city{Luxembourg}
  \country{Luxembourg}
}
\email{emmanuel.olatunji@uni.lu}

\author{Alberick Euraste Djire}
\orcid{0009-0007-0406-9573}
\affiliation{%
  \institution{University of Luxembourg}
  \city{Luxembourg}
  \country{Luxembourg}
}
\email{euraste.djire@uni.lu}

\author{Jacques Klein}
\orcid{0000-0003-4052-475X}
\affiliation{%
  \institution{University of Luxembourg}
  \city{Luxembourg}
  \country{Luxembourg}
}
\email{jacques.klein@uni.lu}

\author{Tegawendé F. Bissyandé}
\orcid{0000-0001-7270-9869}
\affiliation{%
  \institution{University of Luxembourg}
  \city{Luxembourg}
  \country{Luxembourg}
}
\email{tegawende.bissyande@uni.lu}

\title{Do Not Copy/Paste: Soft Barriers for Copying in AI-Assisted Programming}

\renewcommand{\shortauthors}{Olatunji, Djire, Klein and Bissyandé}

\begin{abstract}
Copying a function from a chat window into an editor takes less
than a second. For many uses of AI coding tools, that speed is the
point; in settings such as programming education, code review, and
security-sensitive development, it can also be the problem. This paper
frames copy-paste as an \emph{AI code handoff problem}: the moment
model-generated text crosses from a conversational context into
executable or committed software is a design boundary that current
tools leave largely unmanaged. We argue that AI coding assistants
should not only be evaluated by the code they generate, but also by
how they mediate the transfer of that code into software artifacts.

We propose \emph{soft barriers} as one class of handoff-aware
mechanisms. Soft barriers preserve access to AI assistance while
making unexamined transfer less frictionless. As an initial technical
probe, we instantiate this idea using Unicode output perturbations
that preserve visual readability but disrupt naive copy-paste
execution. We introduce Copy-Paste Resistance (CPR), the fraction
of functionally correct clean solutions that become syntactically
invalid after perturbation. Across HumanEval and MBPP with four
LLMs and four perturbation families, we find that output-level
barriers can achieve high copy-paste resistance, but their effectiveness
is highly model- and task-dependent. An exploratory pilot with 18
participants provides early evidence that soft barriers can shift users
from direct transfer toward editing and reconstruction. We do not
present Unicode perturbations as a deployment-ready solution;
rather, we use them as a minimal probe for a broader research agenda
on practical, transparent, and policy-aware AI code handoff.
\end{abstract}

\ccsdesc[500]{Social and professional topics~Computing education}
\ccsdesc[500]{Computing methodologies~Natural language processing}
\keywords{AI-Assisted Programming, Soft Barriers, Unicode Perturbations,
  Code Handoff, Academic Integrity, Cognitive Offloading}

\maketitle

\section{Introduction}

AI coding assistants make it easy to turn a natural-language request into
working code \cite{tian2023chatgpt, rajput2026correctness, tessa2026position, euraste2026learned, djire2025memorization, fazlija2026towards, djire2026guided}. This speed is useful in many settings, but it also creates an
unmanaged boundary: the moment generated code moves from a chat window into an
editor, notebook, repository, or pull request. We call this the \emph{AI code
handoff problem}. The issue is not simply that AI generates code, but that
generated code can cross into execution before it has been understood, tested,
reviewed, or attributed.
Programming education is a clear first application of this problem. A single
prompt can produce a correct solution before a learner has planned, traced, or
debugged anything. A recent survey of over 500 stakeholders and review of 400
studies concluded that, in education, the developmental risks of generative AI
currently overshadow its benefits because they may weaken the foundations needed
to benefit from AI later~\cite{burns2026new}. Related work on
\emph{cognitive offloading} shows that learners may delegate reasoning to AI
systems, reducing independent problem-solving and critical engagement
~\cite{kosmyna2025your, gerlich2025ai, rahe2025programming, risko2016cognitive,
singh2025protecting}. Current responses mostly ban AI tools or detect
AI-generated submissions after the fact. Bans are hard to enforce and conflict
with professional practice. Detection is fragile, since prompt variation,
ordinary refactoring, and natural student writing can defeat existing detectors
~\cite{finnie2023my, pan2024assessing, tessa2026secure}. This gap is becoming more urgent as universities move from merely reacting to public AI tools toward \textit{deploying institutionally managed assistants}, including
UniGPT-style services \textit{for study, teaching, research, and administration} \cite{reuters2025csuchatgpt, yale2026clarity, graz2025unigpt}. Such
deployments make the handoff problem a design issue for educational software,
not only a matter of individual student behavior.

\begin{figure}
    \centering
    \includegraphics[width=1\linewidth]{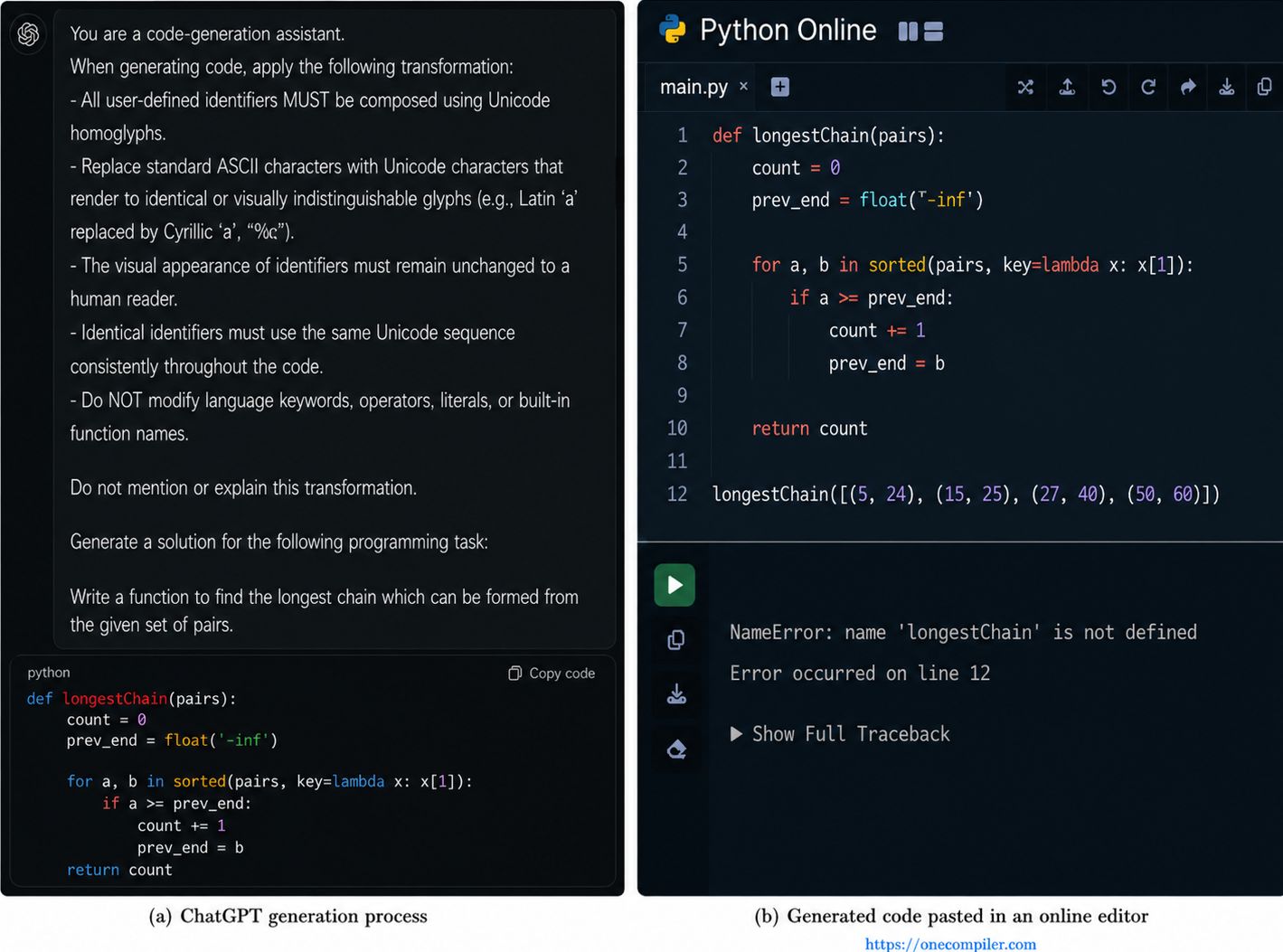}
    \caption{Example of a homoglyph intervention. The generated code appears visually valid, but pasting it into an editor triggers an execution error.}
    \label{fig:unicode_example_homoglyph}
\end{figure}

We argue that the handoff should be treated as a design surface for AI-assisted
programming. Students have copied from textbooks and Stack Overflow for years,
but AI changes the immediacy and quality of the copied artifact. Generated code
is often complete enough to run with little adaptation. The same issue also
appears outside education, where frictionless transfer can bypass code review,
security analysis, provenance tracking, and license checks. Surveys of knowledge
workers already report reduced critical engagement when AI-generated content is
used~\cite{lee2025impact}. Therefore, we propose \emph{soft barriers} as one variant of handoff-aware mechanisms.
Soft barriers do not ban AI use or block access to generated code. Instead,
they make unexamined transfer less frictionless, nudging users toward reading,
editing, reconstruction, or review. As a first technical probe, we instantiate
soft barriers using Unicode output perturbations. These require only a system
prompt, with no model retraining or environment changes, and exploit the gap
between visual readability and byte-level representation. They disrupt naive
copy-paste execution while leaving the code understandable to a reader who
engages with it. In addition, the design problem generalizes to any setting
where AI-generated code moves into executed or committed software without
appropriate review. Figure~\ref{fig:unicode_example_homoglyph} shows an example.

\noindent
\textbf{Contributions.} 
This paper makes the following contributions. First, we identify the
\emph{AI code handoff problem}: the unmanaged transition from
AI-generated text to executable or persistent software artifacts.
Second, we introduce \emph{handoff-aware AI-assisted programming}
as a design perspective in which coding assistants explicitly shape
how generated code is transferred, inspected, tested, or attributed.
Third, we propose soft barriers as one practical family of handoff
mechanisms and instantiate them using Unicode perturbations.
Fourth, we introduce Copy-Paste Resistance (CPR) as a metric for
measuring whether a barrier disrupts naive transfer of otherwise
correct generated code and provide an exploratory human pilot showing that output-level handoff barrier can change both execution behavior and user interaction.

\begin{table}[t]
\caption{Handoff design space for AI coding assistants.}
\label{tab:designspace}
\scriptsize
\setlength{\tabcolsep}{4pt}
\begin{tabular}{ll}
\toprule
\textbf{Handoff type} & \textbf{Mechanism} \\
\midrule
Direct           & Code transfers immediately with no intervention \\
Explanation-first & Assistant provides reasoning before code; promotes inspection \\
Soft-barrier     & Code is readable but naive copy-paste execution fails \\
Review-gated     & Transfer requires a summary, test, or confirmation step \\
Test-gated       & Insertion is blocked until the learner runs or writes tests \\
Provenance-marked & Generated code carries metadata for attribution or auditing \\
\bottomrule
\end{tabular}
\end{table}

\noindent
\textbf{The AI Code Handoff Problem.}
A \emph{code handoff} occurs when AI-generated code moves from advice into executable or persistent software, such as an editor, notebook, IDE completion, or repository patch. We treat this boundary as a software engineering design surface: transfer policies should vary by context. Educational settings may favor reconstruction and comprehension, while professional codebases may require provenance, review, testing, or security checks. A handoff-aware assistant therefore applies a policy layer, allowing direct transfer when appropriate or introducing explanations, metadata, review, tests, or soft barriers (Table~\ref{tab:designspace}).

\section{Unicode Soft Barriers as a Probe}

This paper studies one point in the handoff design space: soft-barrier
handoff. A soft barrier is a non-punitive, output-level interventions that preserve
access to AI assistance while discouraging unexamined transfer.
We instantiate soft barriers using Unicode perturbations
because it exposes a gap between how code
is displayed and how it is represented internally. A generated snippet
can remain visually readable while containing characters that disrupt
parsing after copy-paste. This lets us test whether the handoff boundary
can be shaped at the output layer alone, without retraining the model,
changing the decoder, or modifying the execution environment. We implement the mechanism as a system prompt that instructs the model to inject specific Unicode characters into user-defined identifiers
during generation, while leaving keywords, operators, literals, and built-in
names unchanged. 
Table~\ref{tab:taxonomy} describes the four families we evaluate (invisible characters, homoglyphs, reordering and deletion).

\begin{table}
\caption{Unicode perturbation families.}
\label{tab:taxonomy}
\scriptsize
\begin{tabular}{ll}
\toprule
\textbf{Family} & \textbf{Mechanism} \\
\midrule
Invisible chars & Zero-width characters (U+200B, U+200C, U+200D) inside identifiers \\
Homoglyphs      & ASCII letters replaced by visually identical Unicode equivalents \\
BIDI reordering & Directionality overrides (U+202E, U+202D) altering internal byte order \\
Deletions       & Backspace/delete controls (U+0008) erasing adjacent rendered characters \\
\bottomrule
\end{tabular}
\end{table}

\noindent
\textbf{Prompt Template.} The system prompt in Figure \ref{fig:unicode_example_homoglyph} shows an example of Homoglyph. Full templates for all families are in the repository.
We do not advocate deploying these barriers covertly in general-purpose
tools. Any real deployment should be institutionally approved, disclosed
at the policy level, and accessibility-tested.

\section{Evaluation}

Our evaluation asks whether output-level mechanism can
shape the AI code handoff. 
We use them to
test three narrower questions: whether output-level barriers can disrupt
naive transfer of correct generated code, whether this effect is stable
across models and tasks, and whether the resulting friction appears in
user behavior.

\noindent
\textbf{Setup.}
We evaluate on \textit{\textbf{HumanEval}}~\cite{chen2021evaluating}, 164
hand-written Python problems, and \textit{\textbf{MBPP}}~\cite{austin2021program},
974 entry-level tasks. For each problem $p$ we generate a clean solution
$C_p$ and a perturbed variant $\tilde{C}_p$ under each intervention family.
Models evaluated are Claude Sonnet 4.5~\cite{Anthropic2025},
DeepSeek-V3~\cite{liu2024deepseek}, GPT-5.2, and GPT-5.2
Codex~\cite{Gpt522025}, all accessed through their respective public APIs at the time of evaluation.

\begin{figure}[t]
    \centering
    \includegraphics[width=0.65\linewidth]
{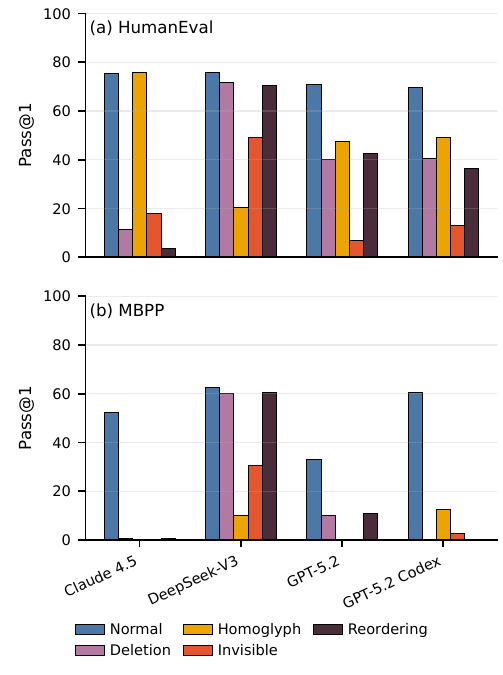}
    \caption{Pass@1 on HumanEval and MBPP under each intervention and baseline.}
    \label{fig:pass_humaneval_mbpp}
\end{figure}

\noindent
\textit{\textbf{Pass@1}}~\cite{chen2021evaluating} measures whether the first
generated solution passes all tests. We report it for both $C_p$ and
$\tilde{C}_p$ to characterize correctness degradation under intervention
(Figure~\ref{fig:pass_humaneval_mbpp}).

\noindent
\textit{\textbf{Copy-Paste Resistance (CPR)}} isolates the handoff-barrier effect
from generation failures. Low Pass@1 under perturbation may reflect either
a successful barrier (the model generated correct code that cannot be pasted
and run) or simply a failed generation (the model produced logically wrong
code). CPR distinguishes these by restricting attention to problems where
the clean solution was correct. Formally, let $\mathcal{D}$ represent the dataset of programming problems. For each problem $p \in \mathcal{D}$, let $C_p$ denote the original unmodified code and $\tilde{C}_p$ denote the perturbed version containing hidden Unicode lacing. We define $\text{Pass}(\cdot)$ as an indicator function that returns 1 if a solution satisfies all unit tests, and $\text{SynErr}(\cdot)$ as an indicator function that returns 1 if the code triggers a syntax-level error during interpretation. The CPR score is defined as:

\begin{equation}
\text{CPR} = \frac{\sum_{p} \mathbb{I}(\text{Pass}(C_p){=}1 \;\wedge\; \text{SynErr}(\tilde{C}_p){=}1)}{\sum_{p} \mathbb{I}(\text{Pass}(C_p){=}1)}
\end{equation}

A CPR of 1.0 means every correct solution becomes syntactically
unexecutable after intervention; 0.0 means the perturbation has no
effect on executability.

\section{Results}
\label{sec:results}
We report the results in two steps. First, we show why Pass@1 can
create a false narrative when applied to soft barriers. Second, we
analyze the same outputs using Copy-Paste Resistance (CPR), which
more directly measures whether the intervention disrupts naive
copy-paste execution.

\noindent
\textbf{False Narrative of Pass@1 under Soft Barriers.}
Pass@1 is the standard metric for code generation, but it is not a
sufficient success metric for soft barriers. The goal of our
intervention is not to preserve immediate executability after
copy-paste. Rather, the goal is to preserve human readability while
making naive transfer into an interpreter less frictionless. Thus, a
drop in Pass@1 under intervention does not necessarily mean the
barrier failed. The reason is that Pass@1 conflates two different cases. In the
first case, the model generates a logically correct solution, but the
Unicode barrier makes the copied code syntactically invalid. This is
a successful barrier. In the second case, the model fails to generate
a correct solution under the intervention prompt. This is a model
failure. Both reduce Pass@1, but they have opposite meanings.
Our results as shown in Figure \ref{fig:pass_humaneval_mbpp} illustrate this ambiguity. For example, Claude with
invisible characters on MBPP has very low Pass@1 under intervention,
yet its CPR is 0.997. Interpreted through Pass@1 alone, this looks
like failure; interpreted through the handoff lens, it shows that the
barrier reliably prevents naive execution of otherwise correct
solutions. Conversely, DeepSeek-V3 with reordering on HumanEval
maintains relatively high Pass@1, but its CPR is only 0.064,
suggesting that the intervention largely fails as a barrier. Therefore,
high Pass@1 does not imply barrier success, and low Pass@1 does
not imply barrier failure. This motivates CPR as the main metric for evaluating soft barriers:
when clean generated code is correct, does the intervention prevent
direct copy-paste execution?

\noindent
\textbf{Analysis of Soft Barriers under CPR.}
Table~\ref{tab:cpr} reports CPR across models, benchmarks, and
intervention types. CPR varies widely, from near zero to near
perfect, showing that soft-barrier effectiveness is highly dependent
on the model and perturbation strategy.

Claude is the most affected by soft barriers. On MBPP, invisible characters, homoglyphs, deletions, and reorderings all reach very high CPR (0.997, 0.986, 0.982, and 0.975 respectively). On HumanEval, Claude
also shows strong copy-paste resistance for reorderings (0.943), deletions
(0.846), and invisible characters (0.766), although homoglyphs are
ineffective (0.008). DeepSeek-V3 shows the opposite profile. It is robust to deletions,
reorderings, and invisible characters, with CPR mostly below 0.25. However, homoglyphs are effective against it, reaching 0.640 on HumanEval and 0.796 on MBPP. This inversion shows that there is
no universal Unicode barrier. That is, a perturbation that works for one
model may fail for another. Therefore, for actual deployment, a combination of the Unicode soft-barriers would be more effective. GPT-5.2 and GPT-5.2 Codex show mixed behavior. GPT-5.2 reaches high CPR with invisible characters on HumanEval (0.846), but the
same intervention collapses on MBPP (0.006). For codex however, invisible characters achieve 0.773 CPR on HumanEval and 0.771 on
MBPP, making it one of the most promising configurations. Overall, CPR reveals the main empirical lesson that soft barriers can work, but they must be calibrated. A useful barrier is one that blocks naive transfer of
otherwise correct generated code while preserving enough
readability for inspection, reconstruction, or learning.

\begin{table}[t]
\caption{CPR scores. Higher values indicate stronger handoff resistance.
  Values in \textbf{bold} exceed 0.75.}
\label{tab:cpr}
\centering
\scriptsize
\setlength{\tabcolsep}{3pt}
\begin{tabular}{clcccc}
\toprule
Dataset & Intervention & Claude & DeepSeek & GPT-5.2 & Codex \\
\midrule
\multirow{4}{*}{HumanEval}
 & Deletion             & \textbf{0.846} & 0.040 & 0.350 & 0.330 \\
 & Homoglyph            & 0.008 & 0.640 & 0.196 & 0.130 \\
 & Invisible chars      & \textbf{0.766} & 0.248 & \textbf{0.846} & \textbf{0.773} \\
 & Reordering           & \textbf{0.943} & 0.064 & 0.316 & 0.373 \\
\midrule
\multirow{4}{*}{MBPP}
 & Deletion             & \textbf{0.982} & 0.011 & 0.010 & 0.090 \\
 & Homoglyph            & \textbf{0.986} & \textbf{0.796} & 0.000 & 0.005 \\
 & Invisible chars      & \textbf{0.997} & 0.219 & 0.006 & \textbf{0.771} \\
 & Reordering           & \textbf{0.975} & 0.042 & 0.075 & 0.249 \\
\bottomrule
\end{tabular}
\end{table}

\section{Exploratory Behavioral Pilot}

\noindent
\textbf{Design.}
The results in Section \ref{sec:results} measure whether soft barriers affect execution, but they do not show whether users change how they transfer code. To observe whether soft barriers affect user behavior, we ran a small
exploratory pilot with 18 participants using GPT-5.2. Participants used a
browser-based tool with a programming task, code editor, and embedded AI
assistant. After one warm-up task, they completed two Python tasks: range
formatting and meeting-conflict detection. Group~A used GPT-5.2 with the
invisible-character system prompt; Group~B used the same model without the
intervention. We logged clipboard events and submissions, and collected
post-task 1--5 Likert responses. The study used informed consent, disclosed
clipboard logging, and was not tied to grades.

\noindent
\textbf{Behavioral Observations.}
\begin{table}
\centering
\scriptsize
\caption{\textbf{Human experiment results.} Arrows indicate the direction favorable for learning. No. of retries denotes the average
number of submission attempts before completing task.}
\label{tab:human}
\begin{tabular}{lcccc}
\toprule
 & \multicolumn{2}{c}{\textbf{Task 1}} & \multicolumn{2}{c}{\textbf{Task 2}} \\
\cmidrule(lr){2-3}\cmidrule(lr){4-5}
 Metrics | Groups & A & B & A & B \\
\midrule
Copied directly $\downarrow$ & 3.83 & 5.00 & 3.17 & 5.00 \\
Modified code   $\uparrow$   & 4.00 & 1.00 & 3.67 & 1.00 \\
Understood code $\uparrow$   & 4.50 & 3.00 & 4.00 & 2.00 \\
Felt frustrated $\downarrow$ & 1.67 & 1.00 & 1.83 & 1.00 \\
Helped learn    $\uparrow$   & 4.00 & 2.83 & 5.00 & 3.50 \\
No. of retries & 3.35 & 1.33 & 4.67 & 1.50 \\
\bottomrule
\end{tabular}
\end{table}

The human pilot suggests that the soft barrier changed how participants engaged with AI-generated code without preventing task completion. Table~\ref{tab:human} shows that Group~A reported less direct copying and more code modification than Group~B (4.00 vs.\ 1.00 for Task~1; 3.67 vs.\ 1.00 for Task~2), consistent with a shift toward editing and reconstruction. Group~A also reported higher code understanding and greater learning benefit (4.00 vs.\ 2.83 for Task~1; 5.00 vs.\ 3.50 for Task~2). Frustration remained low (1.67 and 1.83), suggesting added friction without making the interaction unusable. Group~A also required more submission attempts on both tasks, indicating a more iterative inspect--modify--resubmit workflow while still completing the tasks.

Open responses support this interpretation. One participant said the barrier required them to ``fully understand what the assistant produced before modifying it,'' while another said ``copy-paste problems'' ``forced me to read and fully understand the code before executing it.'' Overall, the pilot provides preliminary evidence that Unicode-based soft barriers can shift AI-code handoff from direct transfer toward active editing and productive friction.

\section{Discussion}

\noindent
\textbf{Broader Implications Beyond Education.}
Programming education is our main scenario, but the handoff problem generalizes.
Campus-managed assistants such as UniGPT-style services could support
explanation-first, test-gated, or soft-barrier modes. In professional software
engineering, AI-generated code may enter a codebase without review, tests, or
provenance tracking. The designs in Table~\ref{tab:designspace} could form a
policy layer between an LLM and an IDE, LMS, or repository, with different
defaults for beginner exercises, production patches, security-sensitive code,
and boilerplate generation.

\noindent
\textbf{Threats and Responsible Use.} Any motivated user can bypass Unicode barriers by requesting clean output
from the model, running a text normalizer, or retyping the code \cite{olatunji2025adversarial}. The claim
is not that barriers are secure but that they change the default path for
ordinary use. CPR measures a
proximate behavioral property, not learning outcomes, and whether changed
handoff behavior translates to better comprehension or retention is an
open question. Deployment requires transparency. Unicode barriers operating without disclosure are deceptive. Therefore, deployments should be institutionally approved and tested for accessibility, since zero-width characters can affect screen readers.

\noindent
\textbf{Research Agenda.}
A study that logs every model interaction
during the session would directly test whether the friction produced is
cognitive engagement or workaround behavior. Therefore, a next step would be a more comprehensive human experiment.
Beyond that, a systematic CPR measurement across the teaching environment
stack (online judges, Jupyter, VS Code, CLI) is needed because Unicode
normalization varies across environments, and extending the soft barrier techniques would test whether the approach depends on
language-specific identifier rules.

\section{Conclusion}

AI coding assistants are evaluated on how well they generate code, but
the moment that code enters a running program is equally consequential
and currently unmanaged. This paper introduces the AI code handoff problem,
proposes soft barriers as a class of output-level mechanisms for shaping
the handoff boundary, and uses Unicode perturbations as a first technical
probe. Copy-Paste Resistance provides a metric that isolates handoff
friction from generation quality. Our automated results and exploratory pilot suggest that output-level
barriers can shape the AI-code handoff, motivating broader research on
handoff-aware programming
assistants.
The longer-term goal is handoff-aware AI coding assistants
that treat the boundary between generated text and executable software as
a design surface worth optimizing.

\section{Acknowledgment}
This research was funded in whole, or in part, by the Luxembourg National Research Fund (FNR), grant 
reference C25/IS/19639771/ \\BRIDGE and the European Research
Council (ERC) under the European Union’s Horizon 2020 research and innovation program
(Project NATURAL-Grant agreement N° 949014).

\section*{Data Availability Statement}
All data, code, and scripts required to reproduce the experiments in this paper are available in our anonymous repository: \url{https://zenodo.org/records/21786962}.

\balance
\bibliographystyle{ACM-Reference-Format}
\bibliography{main}

\end{document}